\documentclass{article}
\title{$f(\az)$ Cosmological solutions with varying speed of light}
\author{A. Izadi and A. Shojai
\\ Department of Physics, University of Tehran,
\\ North Karegar Ave., Tehran, Iran.}
\date{}
\def\az{\mathcal{R}}
\begin{document}
\maketitle
\begin{abstract}
We consider $f(\az)$ modified gravity theories for describing varying speed of light in a spatially flat FRW model, and find some exact solutions. Also we examine the dynamics of this model by dynamical system method assuming a $\Lambda\textrm{CDM}$ background and we find some exact solutions by considering the character of critical points of the theory in both formalisms. The behaviour of the speed of light is obtained. 
\end{abstract}
\section{Introduction}
The late-time accelerated expansion of the universe is an important challenge to the cosmological models. There is an  observational evidence based on Type Ia supernovae standard candles\cite{1} and also on standard rulers\cite{2,3} that the Universe is in a phase of accelerating expansion now.There are  several theoretical approaches\cite{4,5} towards the understanding of this accelerating expansion. The simplest one is to assume the existence of a positive cosmological constant which is small enough to be dominating only at recent times.  The predicted cosmic history (assuming spatial flatness) is then
\begin{equation}\label{H}
(\frac{\dot{a}}{a})^{2}= H_{0}^{2}[\Omega_{0m}(1+z)^{3}+ \Omega_{0r}(1+z)^{4}+ \Omega_{\Lambda}]
\end{equation}
which provides an excellent fit to the  observational data\cite{2}. Models with a cosmological constant suffer from the coincidence problem which is the need for an extreme fine-tuning of the cosmological constant. To address this problem two classes of models have been proposed. In the first class one attributes the accelerating expansion to a dark energy (usually the energy of a scalar field called quintessence) which has repulsive gravitational properties due to its negative pressure\cite{6}. The role of dark energy can also be played by Chaplygin gas\cite{7}, topological defects \cite{8}, holographic dark energy\cite{9}, etc. 

The second class of models look for an accelerating expansion via  modification of general relativity on cosmological scales. Examples of these models include scalar-tensor theories\cite{10,11}, $f(\az)$ modified gravity theories\cite{12}, braneworld models\cite{13}, and so on. These models naturally allow\cite{10,14} for a super accelerating expansion which the effective dark energy equation of state parameter $w=p/\rho$ crosses the phantom divide line $w=-1$. Such a crossing is consistent with some current cosmological data\cite{15}.

The advantage of $f(\az)$ theories of gravity is that no extra degree of freedom is introduced and the accelerating expansion is produced by the Ricci scalar (dark gravity) whose physical origin is well understood. This is in contrast to other models where the origin and physical meaning of the extra degree of freedom is unclear.
It has been shown that for appropriate forms of $f(\az)$ the action can produce accelerating expansion at late times in accordance with SnIa data \cite{16}.
On the other hand, the main drawback of such theories is that they are seriously constrained
by local gravity experiments\cite{18,19,20}.

It can be shown\cite{18} that $f(\az)$ models are dynamically equivalent to scalar--tensor theories with vanishing Brans--Dicke parameter $(\omega =0)$ and a special type of potential.
This implies that in principle the reconstruction of $f(\az)$ from a particular cosmic history $H(z)$ can
be performed in a similar way as in the case of the scalar--tensor theories\cite{10,14}. However, the vanishing of the Brans--Dicke parameter requires some modifications of the reconstruction methods especially when the reconstruction extends through the whole cosmic history through the radiation and matter eras. The dynamical systems approach followed in the present study illustrates these modifications.

The construction of cosmological models incorporating late accelerating expansion based on $f(\az)$ theories has been an issue of interesting debate recently. This debate started from Ref.\cite{23} which
demonstrates that $f(\az)$ theories behaving as a power of $\az$ at large or small $\az$ are not cosmologically viable because they have the wrong expansion rate in the matter dominated era. This conclusion was challenged in Ref.\cite{24} claiming that wide classes of $f(\az)$ gravity models including matter and accelerating
phases can be phenomenologically reconstructed by means of observational data. The debate continued with
the recent work\cite{25} where a detailed and general dynamical analysis of the cosmological evolution of $f(\az)$ theories was performed. It was shown that even though most functional forms of $f(\az)$ are not cosmologically viable due to the absence of the conventional matter era required by data, there are special forms of $f(\az)$ that can be viable with appropriate initial conditions.

There are some ideas suggesting that the constants of nature, such as the speed of light, should be space--time dependent\cite{26,27,28,29}. Theories with varying speed of light (VSL) have been firstly proposed by Moffat, Albrecht, Magueijo and Barrow\cite{29,30} as an alternative approach to the inflation mechanism for solving some problems of Big-Bang cosmological models\cite{29,31}. In their formulation the Lorentz invariance is broken and there is a preferred frame, in which the speed of light depends only on time. In this frame there exists a pre-set function \cite{30,32} representing the speed of light and enters in the Friedman equations as an input.

It is a well-known fact that it is possible to have a varying speed of light theory and preserving the general covariance and local Lorentz invariance\cite{33}. The price that have to be paid for this, is to introduce a time--like coordinate $x^{0}$ which is not necessarily equal to $ct$. In terms of $x^{0}$ and $\vec{x}$, one has local Lorentz invariance and general covariance. The physical time $t$, can only be defined when $dx^0/c$ is integrable.

The most general scalar-tensor action of gravity which allows for a dynamical speed of light is illustrated in \cite{34}. This action is previously analysed by many authors. Demianski et al. \cite{35} present a class of cosmological models derived from N\"oether symmetry requirement. These models describe accelerating evolution of an FRW universe filled with matter and exhibit power law dependence of the coupling factor and the potential to the scalar field. There is also some tracking solutions of this model, in which the time evolution of the scalar field tracks the expansion rate of the universe. 

Here we shall investigate the exact cosmological solutions with varying speed of light in the framework of $f(\az)$ modified gravity theories. In the following sections we shall find some exact cosmological solutions for the spatially flat universe. In section 3 we shall examine the dynamics of this theory by dynamical system method assuming a $\Lambda\textrm{CDM}$ background. Considering the character of the critical points of the theory we find some exact cosmological solutions. In section 4 we shall use the solution of the above system to reconstruct the cosmological evolution and functional form of the function $f(R)$ and the speed of light.
\section{The Model}
The action which we use here is similar to the Jordan-VSL action in \cite{34}, except that we have changed the gravitational part to $f(\az)$:
\begin{equation} \label{action}
S = \frac{1}{16\pi G} \int d^{4}x \sqrt{-g}\left ( h(\psi) f(\az) - 2U(\psi) - Z(\psi) g^{\mu\nu} \partial_{\mu}\psi \partial_{\nu}\psi \right ) + S_{m}[\phi_{i};g_{\mu\nu}]
\end{equation}
Here $h(\psi) = (\frac{c}{c_{0}})^{4}$ and $U(\psi)$ are arbitrary regular functions of the scalar field $\psi$ (the field that generates varying speed of light), representing the coupling of the scalar field $\psi$ with geometry and it's potential energy density respectively. $c_{0}$ is the constant velocity of light and hereafter we shall put $8\pi G = c_{0}^{4}=1$. The first part of the above action functional is the gravitational part, including $f(\az)$ and a dynamical term for the velocity of light with arbitrary coupling function $ Z(\psi)$. The latter is the action of the matter fields, $\phi_{i}$, and we assumed that it does not involve the scalar field $\psi$, so that the matter is minimally coupled to gravity. As emphasized in the introduction, here it is assumed that there is a time-like coordinate $x^{0}$ and since $dx^0/c$ is not necessarily integrable, it is not always possible to define \textit{time}. 
It has to be noted that $Z(\psi)$ can always be set equal to unity by a redefinition of the field $\psi$. 
Finally it has to be noted that here we adopt a metric approach so that the metric, the scalar field $\psi$, and the matter fields $\phi_i$ are dynamical variables. Varying the action with respect to metric and $\psi$ field gives respectively:
\[
h (f'\az_{\mu\nu}-\frac{1}{2}fg_{\mu\nu}) -\nabla_{\mu}\nabla_{\nu} (hf') + g_{\mu\nu} \nabla_{\alpha} \nabla^{\alpha} (hf') =
\]
\begin{equation}
T_{\mu\nu} + \partial_{\mu}\psi \partial_{\nu}\psi - \frac{1}{2}g_{\mu\nu} (\partial_{\alpha}\psi)^{2} - g_{\mu\nu} U
\end{equation}
and
\begin{equation}
\nabla_{\mu}\nabla^{\mu}\psi = \frac{dU}{d\psi} - \frac{1}{2} f\frac{dh}{d\psi}
\end{equation}
where $f'$ is the derivative of $f$ with respect to $\az$.

The weak equivalence principle holds because the matter fields are minimally coupled to the metric. This means that we have $\nabla_{\mu}T^{\mu}_{\nu} = 0$ where the energy momentum tensor of matter is defined as usual; namely $T^{\mu\nu} = \frac{2}{\sqrt{-g}} \frac{\delta S_{m}}{\delta g_{\mu\nu}}$
In a cosmological context, applying the above field equations to the spatially flat FRW universe in which the metric has the following form:
\begin{equation}
ds^{2} = -(dx^0)^2 + a(x^0)^2\left(dr^2 + r^{2} d\Omega^{2}\right)
\end{equation}
and assuming the matter field as a perfect fluid, we have:
\begin{equation}\label{1st friedmann}
3hf'H^{2} = \rho +\frac{1}{2}\dot{\psi}^{2} + \frac{1}{2}\az hf' - \frac{1}{2}hf + U - 3H(\dot{hf'})
\end{equation}
\begin{equation}\label{2nd friedmann}
-2hf' \dot{H} = \rho + p + \dot{\psi}^{2} + ( \ddot{hf'}) - H( \dot{hf'})
\end{equation}
\begin{equation}\label{wave}
\ddot{\psi} + 3H \dot{\psi} = \frac{1}{2} \frac{dh}{d\psi} f - \frac{dU}{d\psi}
\end{equation}
\begin{equation}\label{cons}
\dot{\rho} + 3H( \rho + p) = 0
\end{equation}
where a dot over any quantity denotes derivative with respect to the time-like coordinate $x^{0}$.
These are $c-$variable Friedman equations, the field equation of $\psi$  and the conservation law respectively. $H(x^{0}) = \frac{1}{a} \frac{da}{dx^{0}}$ is the Hubble parameter, $\rho$ and $p$ are the energy and pressure density of a perfect fluid considered as matter field.  These equations form a coupled set of non-linear differential equations for $H(x^{0})$ and $\psi(x^{0})$. The time-like coordinate $x^{0}$ is related to cosmic time by the relation:
\begin{equation}
dt = \frac{dx^{0}}{c} = h^{-\frac{1}{4}}dx^0
\end{equation}  
In the cosmological application $dx^0/c$ is integrable and gives the physical time. Therefore, the physical Hubble parameter $H_{p}(t) = \frac{1}{a} \frac{da}{dt}$ can be evaluated as $H_{p}(t) = H(x^{0})\frac{dx^{0}}{dt}$.
Substituting $\frac{1}{H(x^{0})} \frac{d}{dx^{0}}$ by $\frac{1}{H_{p}(t)} \frac{d}{dt}$ in eq.(\ref{cons}) gives: $\frac{d\rho}{dt} + 3H_{p} (\rho + p) = 0$
This shows that in this model the conservation equation (\ref{cons}) is valid even in terms of the cosmic time.
Since in this model $\az = 6( \dot{H} + 2H^{2})$, we can rewrite the eqs.(\ref{1st friedmann}), (\ref{2nd friedmann}) in these forms:
\[
3hf'H^{2} = \rho + \frac{1}{2} \dot{\psi}^{2} + 3(\dot{H} + 2H^{2})hf' - \frac{1}{2}hf +
\]
\begin{equation}\label{fr1}
U - 3H[ \dot{h}f' + 6hf''( \ddot{H} + 4H \dot{H})]
\end{equation}
\[
-2hf'\dot{H} = \rho + p + \dot{\psi}^{2} + \dot{h}f' + 12 \dot{h}f''( \ddot{H} + 4H \dot{H}) + 36hf'''( \ddot{H} + 4H\dot{H})^{2} + 
\]
\begin{equation}\label{fr2}
6hf''[\stackrel{...}{H} + 4\dot{H}^{2} + 4H \ddot{H}] - H[ \dot{h}f' + 6hf''( \dot{H} + 4H\dot{H})]
\end{equation}

The cosmological solutions of the greatest interest are those for which the time evolution of the Hubble parameter is proportional to the inverse of the cosmic time (corresponding to a power law expansion) or a constant (corresponding to de-Sitter expansion). We can thus distinguish two cases, a $c-$dominated universe ($S_{m}=U=0$), and a $(c-\Lambda)-$dominated universe for which $S_{m}=0$ but $U=\Lambda g$ is not zero. This corresponds to adding a constant ($-2\Lambda$) to $f(\az)$ which is equivalent to 
a cosmological constant.

In order to have explicit solutions, we have to choose the form of $f(\az)$. In \cite{37} the VSL Friedman equations for $f(\az)=\az$ is investigated. But for an $f(\az)$ model with constant speed of light, according to the stability conditions for de-Sitter space, we have to choose this form of $f(\az)$:
\begin{equation}
f(\az) =\az - \frac{\mu ^{4}}{\az} + b \az^{2}
\end{equation}
Then the condition for the existence of a de-Sitter solution is $\az_{0} = \sqrt{3}\mu^{2}$, while the stability condition is satisfied if $b > \frac{1}{3\sqrt{3}\mu^{2}}$. For more details see \cite{38}.
Here in this section we shall use the above form of $f$ for our VSL $f(\az)$ model.
\subsubsection{$c-$dominated universe} 
Putting $S_{m}=U=0$  and $f(\az) =\az - \frac{\mu ^{4}}{\az} + b \az^{2}$ in the equations (\ref{fr1}), (\ref{fr2}), we get two independent equations. Assuming a power-law dependence for the coupling coefficient $h(\psi)$, these equations have the following solution:
\begin{equation}
H \sim constant,\ \ \ \ \ \ \  \psi \sim e^{\alpha x^{0}}, \ \ \ \ \ \ \ h\sim\psi^{2}
\end{equation} 
This is a special choice which is used by many authors \cite{33,39,40}. 
The coupling function $h(\psi)\sim \psi^{2}$ is a particular case emerged by requiring the existence of N\"oether symmetry \cite{35} and $\alpha$ is related to the other constants like $\mu$ and $b$.

The cosmic time and the physical Hubble parameter are:
\begin{equation}
t\sim e^{\frac{-\alpha}{2}x^{0}}, \ \ \ H_{p} \sim \frac{1}{t} \longrightarrow a \sim t^{\nu}
\end{equation}
The conditions which one should impose on VSL models are usually inspired by the cosmological puzzles. In order to solve the horizon problem of the standard cosmology, one should set  $\frac{\ddot{a}}{\dot{a}} - \frac{\dot{c}}{c} >0 $ for the early universe(see \cite{26} ) and also one has $ \dot{a}>0 $. So requesting an expanding universe together with the horizon criteria, one gets the following constraint; $\nu >0$.

\subsubsection{($c-\Lambda)-$dominated universe} 
As mentioned before, this era corresponds to a matter free universe but the potential is non-zero. Assuming this form $U= \Lambda g$ in which $\Lambda$ is a constant and demanding power-law expansion for cosmic scale factor, one can find that the solution is:
\begin{equation}
H \sim constant,\ \ \ \ \ \ \  \psi \sim e^{\alpha x^{0}},\ \ \ \ \ \ \  h\sim \psi^{2}.
\end{equation}
and the cosmic time is:
\begin{equation}
t \sim e^{\frac{-1}{2}\alpha x^{0}}.
\end{equation}
\section{Dynamics of VSL $f(\az)$ Cosmology}
Another way to find out some exact solutions for cosmological models is the dynamical system method \cite{41}. In this method by choosing some appropriate variables, one can convert the field equations of the desired theory to a set of autonomous differential equations. Then the \textit{critical points} of the autonomous system describe interesting exact solutions. Also one can use this method to check the stability of the solutions. Dynamics of a  scalar tensor theory in the Jordan frame using metric approach has been considered in \cite{37, 42}. Also dynamics of $f(\az)$ cosmology (not VSL $f(\az)$ cosmology) has been considered in \cite{43}.
Here we consider a class of VSL theories described by action (\ref{action}). We should notice that the volume element is defined as $dx^{0}d^{3}x$ which is different from the canonical volume element, and in the field equations $H$ is not the physical Hubble parameter and derivatives are with respect to $x^{0}$ coordinate.

In section 2, we chose $h(x^{0})$ such that the corresponding solution for cosmic scale factor was physically interesting. Here we impose a general form for physical Hubble parameter which is related to $\Lambda\textrm{CDM}$ cosmology, given by eq.(\ref{H}).
Also we assume $U(\psi) \sim h(\psi)^{n}$ where $n$ is a constant \cite{44}. The form of $f(\az)$ is not fixed here.

Let's rewrite the VSL Friedman equations (\ref{1st friedmann}), (\ref{2nd friedmann}) as:
\begin{equation}\label{fr11}
3hFH^{2} = \rho_{m} + \rho_{rad} + \frac{1}{2} \dot{\psi}^{2} + \frac{1}{2}\az hF - \frac{1}{2}hf +U -3H \dot{h}F - 3Hh \dot{F}
\end{equation}  
\begin{equation}\label{fr22}
-2hF \dot{H} = \rho_{m} + \frac{4}{3}\rho_{rad} + \dot{\psi}^{2}+ \ddot{h}F + 2 \dot{h}\dot{F} + h \ddot{F}-H \dot{h} F - H h \dot{F}
\end{equation}
where $ F:= \frac{df}{d\az}$ and $\rho_{m}, \rho_{rad}$ represent the matter and radiation energy densities which are conserved according to
\begin{equation}
\dot{\rho}_{m} + 3H \rho_{m} = 0, \ \ \ \dot{\rho}_{rad} + 4H \rho_{rad} = 0.
\end{equation}   
In order to study the cosmological dynamics implied by eqs.(\ref{fr11}), (\ref{fr22}) we express them as an autonomous system of first order differential equations. To achieve this, we first write (\ref{fr11}) in the dimensionless form as
\begin{equation}\label{fr111}
1 = \frac{\rho_{m}}{3hFH^{2}} + \frac{\rho_{rad}}{3hFH^{2}} + \frac{\psi '^{2}}{6hF} + \frac{\az}{6H^{2}} - \frac{f}{6FH^{2}} + \frac{U}{3hFH^{2}} - \frac{h'}{h} - \frac{F'}{F}.
\end{equation}
where $' = \frac{d}{d\ln a}\equiv \frac{d}{dN} = \frac{1}{H} \frac{d}{dt}$ 
We now define the dimensionless variables $x_{1},..., x_{7}$ as
$$x_{1}:= \frac{-F'}{F}, x_{2} := \frac{-h'}{h}, x_{3} := \frac{U}{3hFH^{2}}, x_{4} := \frac{-f}{6FH^{2}},$$  
\begin{equation}\label{01}
x_{5} := \frac{\az}{6H^{2}} = \frac{H'}{H} +2, x_{6} := \frac{\psi '^{2}}{6hF}, x_{7} := \frac{\rho_{rad}}{3hFH^{2}} = \Omega_{rad} 
\end{equation}
As one can see $x_{7}$ is in fact $\Omega_{rad}$ and $x_{1}+x_{2}+x_{3}+x_{4}+x_{5}+x_{6} \equiv \Omega_{DE}$ is associated with the curvature dark energy (dark gravity). Defining  $\Omega_{m} \equiv \frac{\rho_{m}}{3hFH^{2}}$ we can write eq.(\ref{fr111}) as
\begin{equation}\label{omega}
\Omega_{m} = 1- x_{1}-x_{2}-x_{3}-x_{4}-x_{5}-x_{6}-x_{7}.
\end{equation} 
Using the defined dimensionless variables, we can express eq.(\ref{fr22}) as
\begin{equation}\label{2}
x'_{1} + x'_{2} = -1-3x_{3} - 3x_{4} - x_{5} + 3x_{6} + x_{7}+ x_{1}^{2} + x_{2}^{2} - x_{2} x_{5} + 2x_{1} x_{2} - x_{1} x_{5}.
\end{equation}
Also, differentiating $x_{3},...x_{7}$ with respect to $N$ we have
\begin{equation}\label{3}
x'_{3}= x_{3} [(1-n)x_{2} + x_{1} - 2x_{5} + 4] 
\end{equation} 
\begin{equation}\label{4}
x'_{4} = \frac{x_1 x_5}{m} - x_4 (2 x_5 - x_1 -4 )
\end{equation}
\begin{equation}\label{5}
x'_5 = -\frac{x_1 x_5}{m} - 2x_5 (x_5 - 2)
\end{equation}
\begin{equation}\label{6}
x'_6 = x_6 (-2 - 2x_5 + x_1 + x_2 ) + x_2 (x_4 + nx_3)
\end{equation}
\begin{equation}\label{7}
x'_7 = x_7 ( x_1 + x_2 - 2x_5 )
\end{equation}
where 
\begin{equation}\label{m-def}
m = \frac{F' \az}{f'} = \frac{f_{,\az\az}\az}{f_{,\az}}
\end{equation}
and $_{,\az}$ implies derivative with respect to $\az$.

The autonomous dynamical system (\ref{2}), (\ref{3}), (\ref{4}), (\ref{5}), (\ref{6}) and (\ref{7}) is the general dynamical system that describes the cosmological dynamics of VSL $f(\az)$ theories. Instead of investigating the above autonomous system for various different behaviours of $m(f(\az))$ we eliminate $m$ from the system by assuming a particular form for $H_{p}(N)$ consistent with cosmological observations. Once $x_{5}(N)$ is known we can solve (\ref{5}) for $\frac{x_1 x_5}{m}$ and substituting in (\ref{4}), we find
\begin{equation}\label{4r}
x'_4 = -x'_5 - 2x_5 (x_5 - 2) - x_4 (2x_5 - x_1 -4)
\end{equation}
which along with (\ref{2}), (\ref{3}), (\ref{6}) and (\ref{7}) describes a new dynamical system which is independent of m. On the other hand one can easily verify that $\frac{H'}{H} = \frac{H'_{p}}{H_{p}} + \frac{1}{4} x_2 $, so by substituting this relation in the above equations, we have:
\begin{equation}\label{22}
x'_1 + x'_2 = -3 - 3x_3 -3x_4 - \frac{H'_p}{H_p} ( 1+x_1 +x_2 ) - \frac{9}{4}x_2 + 3x_6 +x_7 + x_1 ^{2} +\frac{3}{4}x_2 ^{2} + x_1 \left ( \frac{7}{4}x_2 -2 \right )
\end{equation}
\begin{equation}\label{33}
x'_3 = x_3 \left [ (\frac{1}{2}-n)x_2 + x_1 -2\frac{H'_p}{H_p}\right ]
\end{equation}
\begin{equation}\label{44}
x'_4 + x'_5 = -2\left ( \frac{H'_p}{H_p} + \frac{1}{4}x_2 +2 \right )\left (\frac{H'_p}{H_p} + \frac{1}{4} x_2 \right ) - x_4 \left (2 \frac{H'_p}{H_p} + \frac{1}{2}x_2 - x_1 \right )
\end{equation}
\begin{equation}\label{66}
x'_6 = x_6 \left ( -6-2 \frac{H'_p}{H_p} + x_1 + \frac{1}{2}x_2 \right ) + x_2 ( x_4 + nx_3)
\end{equation}
\begin{equation}\label{77}
x'_7 = x_7 \left ( x_1 + \frac{1}{2} x_2 -2 \frac{H'_p}{H_p} -4 \right )
\end{equation}
and $\ \ m = \frac{-x_1}{2\frac{H'_p}{H_p} + \frac{1}{2}x_2}$. 
\\The results of our analysis do not rely on the use of any particular form of $x_5 (N)$(i.e. $H_p (z)$). They only require that the universe goes through the radiation era, matter era and acceleration era.
For the sake of definiteness however, we will assume a specific form for $H(z)$ corresponding to a $\Lambda\textrm{CDM}$ cosmology (\ref{H}) which in terms of $N$, takes the form 
\begin{equation}\label{HH}
H_{p}(N)^{2} = H_{0}^{2} [\Omega_{0m} e^{-3N} + \Omega_{0rad}e^{-4N} + \Omega_{\Lambda} ]
\end{equation}
where $N = \ln a = -\ln (1+z)$ and $\Omega_{\Lambda} = 1-\Omega_{0m} - \Omega_{0rad}$. 
It is straightforward to study the dynamics of the system (\ref{22}), (\ref{33}), (\ref{44}), (\ref{66}) and (\ref{77}) by setting $x'_i =0$ to find the critical points and their stability in each one of the three eras. By setting $x'_i =0$, we have
\begin{equation}\label{55}
x_5 = \frac{H'_p}{H_p} + \frac{1}{4} x_2 +2
\end{equation}
These equations describe the cosmological dynamics of the VSL $f(\az)$ theory. 

In general, we should add a further parameter which could be related to $h_{,\psi}$ and $h_{,\psi\psi}$, beside the above $n$. In fact if we do not try to reconstruct the function $h(\psi)$, such a function can be fixed priori and the corresponding parameter would be, for example, $\frac{h_{,\psi}}{h}$. In such a case $H(N)$ would not be fixed as in our reconstruction approach but would have to be determined by the autonomous system.

The critical points are shown in Tables (1), (2) and (3).
The stability analysis of these Tables assumes that $x_{5}= const$ and therefore it is not identical to the full stability analysis where $x_{5}$ would be allowed to vary. The usual stability analysis of the cosmological dynamical systems assumes a particular cosmological model (e.g. a form of $f(\az)$ or $m$ or $n$) and in the context of this physical law, the stability of cosmic histories $H(N)$ is investigated. In this context clearly a stable cosmic history is the one preferred by model. In the reconstruction approach, however, the stability analysis has a very different meaning. Here we do not fix the model $h(\psi)$ and $f(\az)$ (physical law). Here we fix the cosmic history and allow the physical law $f(\az)$ and $h(\psi)$ to vary in order to predict the required cosmic history. Thus our stability analysis concerns the physical law $f(\az)$ and $h(\psi)$ and not the particular cosmic history. The physically interesting quantities are the values of the critical points we find in each era in the context of the $\Lambda\textrm{CDM}$ cosmic history. These tell us the possible physical laws $f(\az)$ and $h(\psi)$  that can reproduce a $\Lambda\textrm{CDM}$ cosmic history.
Some critical points in each era are not stable. This however does not imply that these points are not cosmologically relevant.
These instabilities are not instabilities of the trajectory $H(N)$ (which we keep fixed) but of the forms of $f(\az)$ which is allowed to vary. Thus they are not so relevant physically since in a physical context $f(\az)$ is assumed to be fixed a priori. Calculation of eigenvalues are too long, but in what follows we don't need them. So we will not write them, and thus we will not present them here.

\begin{table}[h]
\begin{center}
\begin{tabular}{|c||c|c|c|c|}
\hline $x_1$ & $0$ & $ \frac{4m}{1-m}$ & $ \frac{8m(n-1)}{-1+2n+m}$ & $m(4-\frac{1}{2}x^*_{2})$ \\
\hline $x_2$ & $8$ & $ \frac{-8m}{1-m}$ & $\frac{8(m+1)}{-1+2n+m}$ & $x^*_{2}$ \\
\hline $x_3$ & $0$ & $0$ & $x^*_{3}$ & $0$ \\
\hline $x_4$ & $3$ & $ \frac{2m}{1-m^{2}}$ & $\frac{-2}{-1+2n+m}$ & $ \frac{-1}{4(1+m)}x^*_{2}$ \\
\hline $x_5$ & $2$ & $ \frac{-2m}{1-m}$ & $\frac{2(m+1)}{-1+2n+m}$ & $\frac{1}{4}x^*_{2}$ \\
\hline $x_6$ & $-12$ & $ \frac{-8m^{2}}{(1-m)(1-m^{2})}$ & $x^{*}_{6}$ & $x^{**}_{6}$ \\
\hline $x_7$ & $0$ & $\frac{-5m^{3}+9m^{2}+3m+1}{(1-m)(1-m^{2})}$ & $0$ & $0$ \\
\hline 
\end{tabular} 
\end{center}
\caption{The critical points of the system in the radiation era.}	
\label{table1}
\end{table}

In Table (1) the critical points of the system for radiation era is given, in which 
$$x^*_{3} =\frac{-3(1-m) + 2n(1-2m)}{n(-1+2n+m)^{2}\left( 12n(1+m) + 9(1-m) - 6n(1-2m) \right)}\times $$
$$\left( (-1+2n+m) \left(n^{2}+(n-3)(n-2)-mn(53-16m^{2} + 8m(1+4n))\right) \right . $$
$$\left . -16n(1+m)^{2}(m-2)(m-3/2)\right) + \frac{2}{n(-1+2n+m)}$$
and
$$x^{*}_{6} =\frac{4(1+m)}{(-1+2n+m)^2\left( 12n(1+m) + 9(1-m) - 6n(1-2m) \right)}\times $$
$$ \left( (-1+2n+m) \left(n^{2}+(n-3)(n-2)-mn(53-16m^{2} + 8m(1+4n))\right) - \right . $$
$$\left .  16n(1+m)^{2}(m-2)(m-3/2)\right)$$
and
$$ x^{**}_{6} = \frac{(1-16m^{2})}{3} + \frac{-2-27m-12m^{2}+16m^{3}}{12(1+m)}x^*_{2} - \frac{1}{12} (m^{2}-\frac{7}{2}m +3)x_{2}^{*2}$$
and $x^*_{2}$ satifies a cubic equation whose solutions are
$$x^*_{2} = \frac{4(-1+2m)}{m-1}$$
$$ x^*_2 = \frac{-12 m^2 + 16 m^3 - 2 - 27m + \sqrt{64 m^4 - 39m^2 -80 m^3 +52 + 148m}}{2 m^3 - 5 m^2 -m+6}$$
$$x^*_2 = \frac{-12 m^2 + 16 m^3 - 2 - 27m - \sqrt{64 m^4 - 39m^2 -80 m^3 +52 + 148m}}{2 m^3 - 5 m^2 -m+6}
$$

\begin{table}[h]
\begin{center}
\begin{tabular}{|c||c|c|c|c|c|}
\hline $x_1$ & $\frac{2m}{1-m}$ & $0$ & $\frac{3m}{1-m}$ & $\frac{6m(-1+n)}{-1+2n+m}$ & $m(3-\frac{1}{2}x^\dagger_{2})$ \\
\hline $x_2$ & $\frac{2(1-3m)}{1-m}$ & $6$ & $\frac{-6m}{1-m}$ & $\frac{6(1+m)}{-1+2n+m}$ & $ x^\dagger_{2}$ \\
\hline $x_3$ & $0$ & $0$ & $0$ & $x^{\dagger}_3$ & $0$ \\
\hline $x_4$ & $\frac{1-2m}{1-m^{2}}$ & $0$ & $\frac{-1+4m}{2(1-m^{2})}$ & $-\frac{n+2m+1}{2n+2nm+m^{2}-1}$ & $\frac{\frac{3}{2} + \frac{1}{2}x^\dagger_{2} - \frac{1}{8}x_{2}^{\dagger 2}}{(1+m)(-3+\frac{1}{2}x^\dagger_{2})}$ \\
\hline $x_5$ & $\frac{1-2m}{1-m}$ & $2$ & $\frac{1-4m}{2(1-m)}$ & $\frac{n+2m+1}{-1+2n+m}$ & $\frac{1}{4}x^V_{2} + \frac{1}{2}$ \\ 
\hline $x_6$ & $-\frac{6m^{2}-5m+1}{(1-m)(1-m^{2})}$ & $-7$ & $\frac{(1-4m)m}{(1-m)^{2}(1+m)}$ & $x^{\dagger}_6 $ & $x^{\dagger \dagger}_6$ \\
\hline $x_7$ & $-\frac{m(5m^{2}-10m+3}{(1-m)(1-m^{2})}$ & $0$ & $0$ & $0$ & $0$ \\
\hline 
\end{tabular} 
\end{center}
\caption{The critical points of the system in the matter era.}	
\label{table2}
\end{table}

In Table (2) the critical points of the system for the matter era
is given, in which
$$x^{\dagger}_3 = \frac{1}{1+m}\times $$
$$\frac{6n^{2}m^{3} - n^{2}m^{2} - 7mn^{2} + 2n^{2} - 5nm^{3} + 7nm^{2} + 8nm -8n + 3m + 8 - 4m^{2} - m^{3}}{4n^{2} + 4nm - 4n + m^{2} -2m +1}$$
and
$$ x^{\dagger}_6 = -\frac{6n^{2}m^{2} + 5mn^{2} - 2n^{2} +nm^{2} + 7nm + 8n + 2m+1}{4n^{2} +4nm -4n+m^{2} -2m+1}$$
and
$$x^{\dagger \dagger}_6 = \frac{x^\dagger_{2}\left( \frac{\frac{3}{2} + \frac{1}{2}x^\dagger_{2} - \frac{1}{8}x_{2}^{\dagger 2}}{(1+m)(-3+\frac{1}{2}x^\dagger_{2})}\right)}{(1-m)(3-\frac{1}{2}x^\dagger_{2})}$$
and $x^\dagger_{2}$ satisfies a cubic equation, whose solutions are
$$ x^\dagger_2 = \frac{6m}{m-1}$$
$$ x^\dagger_2 = \frac{2\left( 6 m^3 - 8 m^2 -5m +6 + \sqrt{49 m^4 - 76 m^3 - 44 m^2 +72m}\right ) }{-5 m^2 + 2 m^3 +6-m}$$
$$x^\dagger_2 = \frac{2\left ( 6 m^3 - 8 m^2 -5m +6 - \sqrt{49 m^4 - 76 m^3 - 44 m^2 +72m}\right ) }{-5 m^2 + 2 m^3 +6-m}
$$

\begin{table}[h]
\begin{center}
\begin{tabular}{|c||c|c|c|c|}
\hline $x_1$ & $0$ & $\frac{-4m}{1-m}$ & $\frac{-1}{2}mx^\natural_{2}$ & $0$ \\ 
\hline $x_2$ & $0$ & $\frac{8}{1-m}$ & $x^\natural_2$ & $0$ \\ 
\hline $x_3$ & $0$ & $0$ & $0$ & $\frac{1-m}{1+m}$ \\ 
\hline $x_4$ & $-1$ & $\frac{-4+2m}{1-m^{2}}$ & $\frac{-1}{(1+m)}(2+ \frac{1}{4}x^\natural_{2})$ & $ \frac{-2}{1+m}$ \\ 
\hline $x_5$ & $2$ & $\frac{4-2m}{1-m}$ & $\frac{1}{4}x^\natural_{2}+2$ & $2$ \\ 
\hline $x_6$ & $0$ & $\frac{4(-4+2m)}{(1-m)(1-m^{2})}$ & $\frac{x^\natural_{2}(2+\frac{1}{4}x^\natural_{2})}{(1+m)(-6 + \frac{1}{2}(1-m)x^\natural_{2})}$ & $0$ \\ 
\hline $x_7$ & $0$ & $\frac{9(1-m) + m^{2}(13-5m)}{(1-m)(1-m^{2})}$ & $0$ & $0$ \\ 
\hline 
\end{tabular} 
\end{center}
\caption{The critical points of the system in the de-Sitter era.}	
\label{table3}
\end{table}

Finally the critical points of $\Lambda-$ dominated era is given in Table 3, where  
$x^\natural_{2}$ satisfies a cubic equation, whose solutions are
$$ x^\natural_{2} = \frac{-6}{m-1}$$
$$ x^\natural_{2} = -\frac{2\left (5m^{2} - 7m - 9 - \sqrt{49m^{4} - 154 m^{3} + 7m^{2} + 210m +9}\right )}{-5m^{2} + 2m^{3} -m +6}$$
$$x^\natural_{2} = -\frac{2\left (5m^{2} - 7m - 9 + \sqrt{49m^{4} - 154 m^{3} + 7m^{2} + 210m +9}\right )}{-5m^{2} + 2m^{3} -m +6}
$$
\section{Reconstruction of $f(\az)$, $h(\psi)$, $U(\psi)$ }
We can now reconstruct the form of the function $f(R)$ and $h(\psi)$ and $U(\psi)$ that correspond to each one of the critical points of the system shown in tables (1), (2) and (3). These reconstructions are effectively an approximation of these functions in the neighbourhood of each critical point.

Consider a critical point of the form $(\bar{x}_{1},\bar{x}_{2}, \bar{x}_{3}, \bar{x}_{4}, \bar{x}_{5}, \bar{x}_{6}, \bar{x}_{7})$. 
Now we want to reconstruct the form of the potential $U(\psi)$ and $h(\psi)$. Using (\ref{01}), we find that:
\begin{equation}\label{g}
F(N) = F_{0} e^{-\bar{x}_{1}N}, \ \ h(N) = h_{0} e^{-\bar{x}_{2}N} 
\end{equation} 
where $F_0$ and $h_{0}$ are the present value of $F$ and $h$. Assuming that at the present time the velocity of light is equal to $c_0$ we have to set $h_0=1$. Using eq. (\ref{01}) we find 
\begin{equation}\label{p}
\psi(N) = -2(6F_{0})^{\frac{1}{2}} \frac{\bar{x}_{6}^{\frac{1}{2}}}{\bar{x}_{1}+\bar{x}_{2}} e^{-(\bar{x}_{1}+ \bar{x}_{2}) \frac{N}{2}}  + C
\end{equation}
where $C$ is a constant which can be put equal to zero by a shift in $\psi$-field.
Equations (\ref{g}) and (\ref{p}) allow us to eliminate $N$ in favour of $\psi$

\begin{equation}\label{gg}
h(\psi) = \xi  \psi^{\frac{2\bar{x}_{2}}{(\bar{x}_{1} + \bar{x}_{2})}}
\end{equation}
where $\xi= \left ( \frac{1}{24F_{0}} \frac{(\bar{x}_{1} + \bar{x}_{2})^{2}}{\bar{x}_{6}}\right )^{\frac{\bar{x}_{2}}{(\bar{x}_{1} + \bar{x}_{2})}}$. 
\\It is interesting to note that for all solutions obtained here from the analysis of the critical points and for any epoch, the $h$ function has a unique form. Since $c/c_0=h^{1/4}$ we have:
\begin{equation}
\frac{c}{c_0}=a^{-\bar{x}_2/4}
\end{equation} 
Depending on the value of $\bar{x}_2$ given in tables (1), (2) and (3), this can lead to a constant, decreasing or increasing speed of light with respect to the scale factor. Using the horizon criteria one can obtain some restrictions on $\bar{x}_2$ and thus on $m$. For radiation era this condition leads to $\bar{x}_2>4$. For matter era we have $\bar{x}_2>2$. And finally for de-Sitter era it is $\bar{x}_2>-4$. This can be used as a selection rule for model parameters $m$ and $n$.

In a similar way we can reconstruct $U(\psi)$. From eq.(\ref{01}), we have
\begin{equation}\label{u}
U(N) = 3\bar{x}_{3}h(N) F(N) H(x^{0})^{2}= 3\bar{x}_{3}h(N)^{\frac{1}{2}} F(N) H_{p}(N)^{2}
\end{equation} 
Using now the input form of $H_{p}(N)$ (eq.(\ref{HH})), we find the dominant term of $H_{p}(N)$ in each era. 
Using eq.(\ref{p}) and (\ref{u}) we have:
\begin{equation}\label{u-def}
U(\psi) = \lambda \psi ^{\frac{2n\bar{x}_2}{\bar{x}_1 + \bar{x}_2}}
\end{equation}
where
$$\lambda = 3\bar{x}_{3}F_{0}\Omega_{0r}\xi^{\frac{4+\bar{x}_{1} + \frac{1}{2}\bar{x}_{2}}{\bar{x}_{2}}},\ \ \  \textit{Rad. era} $$
$$\ \ \ \ \ = 3\bar{x}_{3}F_{0}\Omega_{0m} \xi^{\frac{3+\bar{x}_{1} + \frac{1}{2}\bar{x}_{2}}{\bar{x}_{2}}},\ \ \ \textit{Mat. era} $$
$$\ \ \ \ \ \ \ \ \ \ \ = 3\bar{x}_{3}F_{0}\xi^{\frac{1}{2}+ \frac{\bar{x}_{1}}{\bar{x}_{2}}} (1 - \Omega_{0r} - \Omega_{0m}),\ \ \  \textit{d.S. era}$$

We can now reconstruct the form of the function $f(\az)$ corresponding  to each one of the critical points of the system. This reconstruction is effectively an approximation of $f(\az)$ in the neighbourhood of each critical point. It is particularly useful because most of the dynamical evolution takes place close to fixed points. 
Using the relation 
\begin{equation}\label{R(N)}
\az = 6(2H^{2} + H' H) = 6\left ( 2H_{p}^{2}h^{-\frac{1}{2}} + H'_{p} H_{p} h^{-\frac{1}{2}} - \frac{1}{4} H_{p}^{2}h' h^{-\frac{3}{2}}\right )
\end{equation} 
In terms of $H_p$ in each era, we obtain
$$\az(N) = \frac{3}{2}\bar{x}_{2}\Omega_{0r} e^{(\frac{\bar{x}_{2}}{2}-4)N}, \ \ \ \textit{Rad. era}$$
$$ = 3\left (1 +\frac{\bar{x}_2}{2} \right )\Omega_{0m} e^{(\frac{\bar{x}_{2}}{2}-3)N}, \ \ \ \textit{Mat. era}$$
\begin{equation}
\ \ = 6  (2+ \frac{\bar{x}_{2}}{4} )(1- \Omega_{0r} - \Omega_{0m})e^{\frac{1}{2}\bar{x}_{2}N}, \ \ \  \textit{d.S. era}
\end{equation}
which leads to
$$ F(\az) = F_{0} \left( \frac{2\az}{3 \bar{x}_{2}\Omega_{0r}}\right)^{\frac{-\bar{x}_{1}}{-4+\frac{1}{2} \bar{x}_{2}}}, \ \ \textit{Rad. era}$$
$$ \ \ \ = F_{0} \left( \frac{\az}{3(1 +\frac{\bar{x}_{2}}{2} )\Omega_{0m}}\right)^{ \frac{-\bar{x}_{1}}{-3+\frac{1}{2} \bar{x}_{2}}},\textit{Mat. era}$$
\begin{equation}
\ \ \ =  F_{0} \left( \frac{\az}{6( 2 +\frac{\bar{x}_{2}}{4} )(1-\Omega_{0r} - \Omega_{0m})}\right)^{\frac{-2\bar{x}_{1}}{\bar{x}_{2}}}, \ \ \textit{d.S. era}
\end{equation}
and by integration we get
$$ f(\az) \sim \az^{\frac{-\bar{x}_{1}}{-4+\frac{1}{2} \bar{x}_{2}} +1}, \ \ \ \textit{Rad. era} $$
$$ \ \ \ \ \ \az^{\frac{-\bar{x}_{1}}{-3+\frac{1}{2} \bar{x}_{2}} +1}, \ \ \  \textit{Mat. era} $$
\begin{equation}
\ \ \ \ \ \ \az^{\frac{-\bar{x}_{1}}{\frac{1}{2} \bar{x}_{2}} +1}, \ \ \ \ \  \textit{d.S. era }
\end{equation}

Although we can reconstruct $f(\az)$ for any critical point and for each one of the three epochs in terms of the values of $\bar{x}_1$ and $\bar{x}_2$, it has to be noted that all of these can be written as
\begin{equation}
f(\az)\sim \az^{1+m}
\end{equation}
as it is clear from the definition of $m$ in equation (\ref{m-def}).
\section{Conclusion}
Here we have investigated analytically the behaviour of VSL $f(\az)$ gravity. We saw that it is possible to produce the background
expansion history $H(z)$ indicated by observations.
Choosing the form $f(\az)\sim\az^{1+m}$ we get $c\sim a^{-\bar{x}_2/4}$. The horizon criteria puts some limitation on $\bar{x}_2$ and thus on $m$. It has to be noted that the form of the potential of the VSL field has to be fixed according to equation (\ref{u-def}).


\begin{thebibliography} {99}
\bibitem{1} S. Perlmutter et al., Astrophys. J. 517, 565 (1999);
J. L. Tonry et al., Astrophys. J. 594, 1 (2003); R. A. Knop
et al., Astrophys. J. 598, 102 (2003); P. Astier et al.,
Astron. Astrophys. 447, 31 (2006).
\bibitem{2} D. N. Spergel et al., Astrophys. J. Suppl. Ser. 148, 175 (2003)
\bibitem{3}  C. Blake, D. Parkinson, B. Bassett, K. Glazebrook, M. Kunz,
and R. C. Nichol, Mon. Not. R. Astron. Soc. 365, 255 (2006).
\bibitem{4} E. J. Copeland, M. Sami, and S. Tsujikawa, Int. J. Mod. Phys. D 15, 1753 (2006).
\bibitem{5} V. Sahni and A. A. Starobinsky, Int. J. Mod. Phys. D 9, 373
(2000); S. M. Carroll, Living Rev. Relativity 4, 1 (2001);
Uzan, Gen. Relativ. Gravit. 39, 307 (2007).
\bibitem{6} C. Wetterich, Nucl. Phys. B302, 668 (1988);  E. J. Copeland, A. R. Liddle, and D. Wands, Ann.
N.Y. Acad. Sci. 688, 647 (1993);  P. G. Ferreira and M. Joyce,
Phys. Rev. Lett. 79, 4740 (1997);  L. Perivolaropoulos, Phys. Rev. D 71,063503 (2005).
\bibitem{7} N. Bilic, G. B. Tupper, and R. D. Viollier, Phys. Lett. B
535, 17 (2002); M. C. Bento, O. Bertolami, and A. A. Sen,
Phys. Rev. D 66, 043507 (2002).
\bibitem{8} A. Friedland, H. Murayama, and M. Perelstein, Phys. Rev. D 67, 043519 (2003).
\bibitem{9} S. Nojiri and S. D. Odintsov, Gen. Relativ. Gravit. 38, 1285 (2006);
M. R. Setare, Phys. Lett. B 644, 99 (2007).
\bibitem{10} B. Boisseau, G. Esposito-Fare`se, D. Polarski, and A. A.
Starobinsky, Phys. Rev. Lett. 85, 2236 (2000); G.
Esposito-Fare`se and D. Polarski, Phys. Rev. D 63,
063504 (2001).
\bibitem{11} J. P. Uzan, Phys. Rev. D 59, 123510 (1999); S. Capozziello,
S. Nojiri, and S. D. Odintsov, Phys. Lett. B 634, 93 (2006);
\bibitem{12} S. M. Carroll, A. De Felice, V. Duvvuri,
D. A. Easson, M. Trodden, and M. S. Turner, Phys. Rev.
D 71, 063513 (2005); M. C. B. Abdalla, S. Nojiri, and S. D. Odintsov,
Int. J. Geom. Methods Mod. Phys. 4, 115 (2007); 
T. P. Sotiriou and S. Liberati, Ann. Phys. (N.Y.) 322, 935
(2007); V. Faraoni, Phys. Rev. D 74, 104017 (2006);
\bibitem{13} 
R. Maartens, Living Rev. Relativity 7, 7 (2004); V. Sahni
and Y. Shtanov, J. Cosmol. Astropart. Phys. 11 (2003) 014;
 C. Bogdanos, A. Dimitriadis, and K. Tamvakis, Phys. Rev. D 75, 087303(2007).
\bibitem{14} L. Perivolaropoulos, J. Cosmol. Astropart. Phys. 10 (2005)
001; S. Nesseris and L. Perivolaropoulos, Phys. Rev. D 75,
023517 (2007); J. Martin, C. Schimd, and J. P. Uzan, Phys.
Rev. Lett. 96, 061303 (2006).
\bibitem{15} S. Nesseris and L. Perivolaropoulos, J. Cosmol. Astropart.
Phys. 01 (2007) 018.
\bibitem{16}
S. M. Carroll, V. Duvvuri, M. Trodden, and M. S. Turner,
Phys. Rev. D 70, 043528 (2004); S. Capozziello, S. Carloni, V. F. Cardone, and A. Troisi,
Int. J. Mod. Phys. D 12, 1969 (2003).
\bibitem{18} T. Chiba, Phys. Lett. B 575, 1 (2003).
\bibitem{19} A. D. Dolgov and M. Kawasaki, Phys. Lett. B 573, 1 (2003).
\bibitem{20} V. Faraoni, Phys. Rev. D 74, 023529 (2006); G. J. Olmo,
Phys. Rev. Lett. 95, 261102 (2005).
\bibitem{23} L. Amendola, D. Polarski, and S. Tsujikawa, Phys. Rev.
Lett. 98, 131302 (2007).
\bibitem{24} S. Capozziello, S. Nojiri, S. D. Odintsov, and A. Troisi,
Phys. Lett. B 639, 135 (2006); S. Nojiri and
S. D. Odintsov, J. Phys. A 40, 6725 (2007).
\bibitem{25} L. Amendola, R. Gannouji, D. Polarski, and S. Tsujikawa,
Phys. Rev. D 75, 083504 (2007).
\bibitem{26} J. Magueijo, Rep. Prig. Phys. 66(2003)2025; G. F. R. Ellis, J.P. Usn, Am.J.Phys. 73 (3)(2005)240; G. F. R. Ellis, Gen. Relativ. Gravit, 39(2007) 511; J. Magueijo, J.Moffat, Gen. Relativ. Gravit. 40(2008)1797.
\bibitem{27} C. Brans, R.H. Dicke, Phys.Rev. 124(1961) 925; R.H. Dicke, Phys. Rev 125(1962) 2163.
\bibitem{28} J. D. Bekenstein, Phys. Rev. D 66 (2002) 123514.
\bibitem{29} J. W. Moffat, Int. J. Mod. Phys. D 2(1993) 351; A. Albercht, J. Magueijo, Phys. Rev. D 59(1999)043516.
\bibitem{30} J. D. Barrow, Phys. Rev. D 59 (1999) 043515.
\bibitem{31}  J. Magueijo, K. Baskerville, Big Bang Riddles and Their Revelations, Cambridge Univ. Press, Cambridge, 2000.
\bibitem{32} J. D. Barrow, J. Magueijo, Phys. Lett. B 447(1999) 246.
\bibitem{33} J. Magueijo, Phys. Rev. D 62(2000) 103521.
\bibitem{34} G. Esposito-Farese, D. Polarski, Phys. Rev. D 63(2001) 063504;\\
A. Riazuelo, J. P. Uzan, Phys. Rev. D 66 (2002) 023525.
\bibitem{35} M. Demianski, E, Piedipalumbo, C. Rubano, P. Scudellaro, Astron. Astrophys. 481 (2) (2008) 279.
\bibitem{37} M. Roshan, M. Nouri, F. Shojai, Phys. lett. B 672(2009) 197-202.
\bibitem{38} T. P. Sotiriou, V. Faraoni, arXiv:0805.1726v2 (2008).
\bibitem{39} F. Shojai, S. Molladavoudi, Gen. Relativ. Gravit. 39 (6) (2007) 795.
\bibitem{40} P.Pedram, S. Jalalzadeh, Phys. Lett. B 660 (2008) 1. 
\bibitem{41} M. Goliath, G. F. R. Ellis, Phys. Rev. D 60 (1999) 023502.
\bibitem{42} S. Capozziello, S. Nesseris, L. Perivolaropoulos, JCAP 0712 (2007) 009.
\bibitem{43} S. Fay, S. Nesseris, L. Perivolaropoulos, Phys, Rev. D. 76, 063504 (2007).
\bibitem{44}  L. Amendola, D. Bellisai, F. Occhionero, Phys. Rev. D 47 (1993) 4267;
               L. Amendola, Phys. Rev. D 60 (1999) 043501.
\end{thebibliography}
\end{document}